\documentclass[journal]{IEEEtran}
\usepackage{mathrsfs}
\usepackage{tikz}
\usepackage{color}
\usepackage{amsmath}
\usepackage{amssymb}
\usepackage{amsthm}
\usepackage{graphicx}
\usepackage{extarrows}
\usepackage{times}
\usepackage{graphics,color}

\newtheorem{proposition}{Proposition}

\newcommand{\onetom}{1,\cdots,m}
\newcommand{\oneton}{1,\cdots,n}

\graphicspath{{figs/}}

\newtheorem{mythm}{Theorem}
\newtheorem{mycol}{Corollary}
\newtheorem{mylem}{Lemma}

\newtheorem{myrem}{Remark}
\begin{document}
\title{
Synchronization, Consensus of Complex Networks and their relationships}

\author{Tianping~Chen,~\IEEEmembership{Senior~Member,~IEEE}
\thanks{This work is  supported by the National Natural Sciences Foundation of China under Grant Nos. 61273211.}
\thanks{T. Chen was with the School of Computer
Sciences/Mathematics, Fudan University, Shanghai 200433, China (tchen@fudan.edu.cn).
}}

\maketitle

\begin{abstract}
In this paper, we focus on the topic Synchronization and consensus of Complex Networks and their relationships. It is revealed that two topics are closely relating to each other and all results given in \cite{Li} can be obtained by the results in \cite{Lu2006}.

\end{abstract}


\section{Introduction}
In recent decades, the synchronization problem of multiagent
systems has received compelling attention from various scientific
communities due to its broad applications. Many natural and synthetic systems, such as neural systems,
social systems, WWW, food webs, electrical power grids,
can all be described by complex networks. In such a network, every node
represents an individual element of the system, while edges
represent relations between nodes. For decades, complex networks have been
focused on by scientists from various fields, for instance,
sociology, biology, mathematics and physics.

In the pioneer work \cite{Pe} (also see \cite{Wang1}), the authors proposed a master stability function near a trajectory, by which local synchronization was investigated. In \cite{Wu}, a distance between node state and synchronization manifold was introduced and global synchronization was discussed.

In \cite{Lu2006}, a general framework is presented to analyze
synchronization stability of Linearly Coupled Ordinary
Differential Equations (LCODEs). The uncoupled dynamical behavior
at each node is general, which can be chaotic or others; the
coupling configuration is also general, without assuming the
coupling matrix be symmetric or irreducible. It was revealed that the left and right eigenvectors corresponding to
eigenvalue zero of the coupling matrix play key roles in the
stability analysis of the synchronization manifold. Different from previous papers, a non-orthogonal projection on the synchronization manifold was first introduced. With this projection, a new approach to investigate the stability of the synchronization manifold of coupled oscillators was
proposed. Novel master stability function near the projection was proposed.

It is clear that linearly Coupled linear system as well as consensus are special cases of the linearly Coupled Ordinary Differential Equations (LCODEs), which are also hot topics in decades. For example, the synchronization of observer based linear systems (see \cite{S}, \cite{T}, \cite{Li} and others).

\section{Unified model and general approach}

In this section,  we present some definitions, denotations and
lemmas required throughout the paper.

In \cite{Lu2006}, following model was discussed
\begin{align}\label{model1}
\frac{d x^{i}(t)}{dt}=f(x^{i}(t),t)
+c\sum\limits_{j=1}^{N}l_{ij}\Gamma x^{j}(t),\quad i=1,\cdots,N
\end{align}
where $x^{i}(t)\in R^{n}$ is the state variable of the $i-th$
node, $t\in [0,+\infty)$ is a continuous time,
$f:R\times[0,+\infty)\rightarrow R^{n}$ is continuous map,
$L=(l_{ij})\in R^{N\times N}$ is the coupling matrix with zero-sum
rows and $l_{ij}\ge 0$, for  $i\ne j$,  which is determined by the
topological structure of the LCODEs, and $\Gamma\in R^{n\times n}$ is an inner coupling matrix. Some time, picking
$\Gamma=diag\{\gamma_{1},\gamma_{2},\cdots,\gamma_{n}\}$ with
$\gamma_{i}\ge 0$, for $i=\oneton$.

\begin{align}\label{model2}
\frac{d x^{i}(t)}{dt}=Ax^{i}(t)
+c\sum\limits_{j=1}^{N}l_{ij}\Gamma x^{j}(t),\quad i=1,\cdots,N
\end{align}
where $A\in R^{n\times n}$.

When $n=1$, $A=0$, we get the following consensus model
\begin{align}\label{model3}
\frac{d x^{i}(t)}{dt}=
\sum\limits_{j=1}^{N}l_{ij} x^{j}(t),\quad i=1,\cdots,N
\end{align}

In case that the state variables $x_{i}(t)$ are not observed. Then, instead of coupling $x_{i}(t)$ (because they are not available), the authors coupled the measured output 
\begin{align*}
\dot{\zeta}_{i}(t)=\sum_{j=1}^{N}l_{ij}y_{i}(t)
\end{align*}
and following observer based synchronization model
\begin{align}\label{model4}
\frac{d x^{i}(t)}{dt}=Ax^{i}(t)
+c\sum\limits_{j=1}^{N}l_{ij}FC x^{j}(t),\quad i=1,\cdots,N
\end{align}
is proposed, where $y(t)=Cx(t)$ is observer measurement $C\in R^{q\times n}$, and $C\in R^{n\times q}$,  was discussed in \cite{Li} and \cite{S,T}.

In \cite{S,T}, the model was written as
\begin{eqnarray}\label{relative1}
\frac{d x^{i}(t)}{dt}=Ax^{i}(t)
+BL\sum\limits_{j=1}^{N}c\mathcal{L}_{ij} x^{j}(t),\quad i=1,\cdots,N,
\end{eqnarray}
where $L\in R^{p\times n}$, and $B\in R^{N\times p}$. It is a special case when the relative states between neighboring agents are available.

It is clear that all these model are special cases of the most general and universal model (\ref{model1}). Therefore, the results given in \cite{Lu2006} can apply to these special cases.

First, we give some basic concepts and necessary background knowledge.

Following Lemma can be found in \cite{Lu2006} (see Lemma 1 in \cite{Lu2006}).
\begin{mylem}\quad 
If $L$ is a coupling matrix with Rank(L)=N-1, then the following items are valid:
\begin{enumerate}
\item If $\lambda$ is an eigenvalue of $L$ and $\lambda\neq 0$,
then  $Re(\lambda)<0$;

\item $L$ has an eigenvalue $0$ with
multiplicity 1 and the right eigenvector $[1,1,\dots,1]^{\top}$;

\item  Suppose $\xi=[\xi_{1},\xi_{2},\cdots,\xi_{m}]^{\top}\in
R^{m}$ (without loss of generality, assume
$\sum\limits_{i=1}^{m}\xi_{i}=1$) is the left eigenvector of $A$
corresponding to eigenvalue $0$. Then,  $\xi_{i}\ge 0$ holds for
all $i=\onetom$; more precisely,

\item $L$ is irreducible if and only if  $\xi_{i}>0$ holds for all
$i=\onetom$;

\item $L$ is reducible if and only if for some $i$, $\xi_{i}=0$.
In such case, by suitable rearrangement, assume that
$\xi^{\top}=[\xi_{+}^{\top},\xi_{0}^{\top}]$, where
$\xi_{+}=[\xi_{1},\xi_{2},\cdots,\xi_{p}]^{\top}\in R^{p}$, with
all $\xi_{i}>0$, $i=1,\cdots,p$, and
$\xi_{0}=[\xi_{p+1},\xi_{p+2},\cdots,\xi_{N}]^{\top}\in R^{N-p}$
with all $\xi_{j}=0$, $p+1\le j\le N$. Then, $L$ can be rewritten
as $\left[\begin{array}{cc}L_{11}& L_{12}\\L_{21}&
L_{22}\end{array}\right]$  where $L_{11}\in R^{p,p}$ is
irreducible and $L_{12}=0$.
\end{enumerate}
\end{mylem}

By definition, any reducible coupling matrix can be rewritten as (see \cite{Lu2004})
\begin{eqnarray*}
L=\left[\begin{array}{cc}
L_{11}&0\\
L_{21}&L_{22}\end{array}\right]
\end{eqnarray*}
or more generally, (see \cite{Lu2006})
\begin{eqnarray*}
L=\left[\begin{array}{cccc}L_{11}&0&\cdots&0\\
L_{21}&L_{22}&\cdots&0\\
\vdots&\ddots&\vdots&\vdots\\
L_{p1}&L_{p2}&\cdots&L_{pp}\end{array}\right]
\end{eqnarray*}
where and for each $q=2,\cdots,p$, $L_{qq}\in
R^{m_{q},m_{q}}$, is irreducible.

\begin{myrem} (see \cite{Lu2006})
In fact, if  $L=[l_{ij}]\in R^{N\times N}$ is a coupling matrix, then $-L$ is a singular M-matrix. Thus,
\begin{itemize}
\item
If $\lambda$ is an eigenvalue of $L$
then  $Re(\lambda)\le 0$;
\item
$L$ has a spanning tree with root $L_{11}$, if and only if
for any $q=2,\cdots,p$, there is some $\bar{q}<q$, such that $L_{\bar{q}q}\ne 0$;
\item
By M-matrix theory, for any $q=2,\cdots,p$, $L_{qq}$ is a non-singular matrix, if and only if there is some $\bar{q}<q$, such that $L_{\bar{q}q}\ne 0$. Equivalently, $L_{qq}$ is a singular matrix (a coupling matrix), if and only if for all $\bar{q}<q$, $L_{\bar{q}q}= 0$. In this case, $L_{11}$ is not a root and $L$ has no spanning tree;
\item
Therefore, $L$ has an eigenvalue $0$ with multiplicity 1, if and only if $L$ has a spanning tree with root $L_{11}$.
\end{itemize}

It is also clear that the so-called master-slave system is a
special case of this model. The nodes in root are masters and others are slaves. Based on previous settings, there is no difference between strongly connected networks and the networks with spanning trees. Therefore, in the following, we assume the networks are strongly connected.
\end{myrem}

Let $[\xi_{1},\cdots,\xi_{N}]^{T}$ be the left eigen-vector corresponding to the eigenvalue $0$ for the matrix $L=[l_{ij}]$. For the model (\ref{model1}) with directed coupling, a nonorthogonal projection of $x(t)$ on the synchronization manifold S, $\bar{X}(t)=[\bar{x}(t),\cdots,\bar{x}(t)]^{T}$, where $\bar{x}(t)=\sum_{i=1}^{N}\xi_{i}x_{i}(t)$, was first introduced in \cite{Lu2006}. It plays a key role in discussing synchronization problem.  For the orthogonal projection $\bar{x}(t)=\frac{1}{N}\sum_{i=1}^{N}x_{i}(t)$ see \cite{Lu2004}. Based on the projection, synchronization is reduced to proving the distance between all nodes $x_{i}(t)$ and the synchronization state $\delta x_{i}(t)=x_{i}(t)-\bar{x}(t)\rightarrow 0$. And (\ref{model1}) can be rewritten as
\begin{eqnarray}
\frac{d \delta x^{i}(t)}{dt}=D f(\bar{x}(t),t)\delta x^{i}(t)
+\sum\limits_{j=1}^{m}l_{ij}\Gamma \delta x^{j}(t)
\end{eqnarray}

Following theorem and corollary were proved in \cite{Lu2006} (see Theorem 1 and Corollary 1 in \cite{Lu2006}).
\begin{mythm}\quad
Let $\lambda_{2},\lambda_{3},\cdots,\lambda_{l}$ be the non-zero
eigenvalues of the coupling matrix $L$. If all variational
equations
\begin{align}
\frac{d z(t)}{dt}=[D f(\bar{x}(t),t)+c\lambda_{k}\Gamma]z(t),\quad
k=2,3,\cdots,l\label{model1a}
\end{align}
are exponentially stable, then the synchronization manifold
$\mathcal S$ is local exponentially stable for the general synchronization model (\ref{model1}). That is $\delta x_{i}(t)=x_{i}(t)-\bar{x}(t)\rightarrow 0$ exponentially.
\end{mythm}

\begin{myrem}
It is clear that Theorem 1 is based on $L_{\infty}$ norm. Following theorem is based on $L_{2}$ norm.
\end{myrem}

\begin{mythm}\quad Let $\lambda_{k}=\alpha_{k}+j\beta_{k}$, $k=2,\cdots,m$, where $j$
is the imaginary unit, be the eigenvalues of the coupling matrix. If there exist a positive definite matrix $P$ and $\epsilon>0$ such that
\begin{eqnarray}
\bigg\{P(D(t)+c\lambda_{k}\Gamma)\bigg\}^{s}<-\epsilon E_{n},\quad
k=2,3,\cdots,m \label{model1a}
\end{eqnarray}
where $D(t)=(D_{ij}(t))$ denotes the Jacobian matrix $D
f(\bar{x}(t),t)$,  $H^{s}=(H^{*}+H)/2$, $H^{*}$ is Hermite
conjugate of $H$, and $E_{n}\in r^{n\times n}$ is identity matrix,
then the synchronization manifold ${\mathcal S}$ is locally
exponentially stable for the coupled system (\ref{model1}).
\end{mythm}

\subsection{Applications to Consensus}

It is clear that for linear systems, globally stable and locally stable are equivalent. Therefore, applying Theorem 1 to the models (\ref{model2}), (\ref{model3}) and (\ref{model4}), we have

\begin{mycol}
Let $\lambda_{2},\lambda_{3},\cdots,\lambda_{l}$ be the non-zero
eigenvalues of the coupling matrix $L$. If all variational
equations
\begin{align}
\frac{d z(t)}{dt}=[A+c\lambda_{k}\Gamma]z(t),\quad
k=2,3,\cdots,l\label{model2a}
\end{align}
are exponentially stable, then the synchronization manifold
$\mathcal S$ is globally exponentially stable for the models (\ref{model2}), (\ref{model3}) and (\ref{model4}) with $\Gamma=FC$.
\end{mycol}

\begin{mycol}
Let $\lambda_{2},\lambda_{3},\cdots,\lambda_{l}$ be the non-zero
eigenvalues of the coupling matrix $L$. If there exist a positive definite matrix $P$ and $\epsilon>0$ such that
\begin{align}
\bigg\{P(A+c\lambda_{k}\Gamma)\bigg\}^{s}<-\epsilon E_{n},\quad
k=2,3,\cdots,m \label{model1b}
\end{align}
are exponentially stable, then the synchronization manifold
$\mathcal S$ is globally exponentially stable for the models (\ref{model2}), (\ref{model3}) and (\ref{model4}) with $\Gamma=FC$.
\end{mycol}

In case $(A,C)$ is detectable, one can find $F$ constructively.

Because $(A,C)$ is detectable, for a fixed $T$,
\begin{align*}
P=\int_{0}^{T}e^{-A^{T}t}C^{T}Ce^{-At}dt>0
\end{align*}
\begin{align*}
PA+A^{T}P &=-\int_{0}^{T}\frac{d}{dt}[e^{-A^{T}t}C^{T}Ce^{-At}]dt\\
&=C^{T}C-e^{-A^{T}t}C^{T}Ce^{-At}<0
\end{align*}
Therefore, there exists $\epsilon>0$ such that
\begin{align*}
PA+A^{T}P-C^{T}C <-e^{-A^{T}t}C^{T}Ce^{-A^{T}t}<-\epsilon I_{n}
\end{align*}
and pick $F=P^{-1}C^{T}$.
\begin{align}
\bigg\{P(A+c\lambda_{k}P^{-1}C^{T}C)\bigg\}^{s}=PA+A^{T}P-cRe(\lambda_{k})C^{T}C
\end{align}
If for all $k=2,3,\cdots,m$, $cRe(\lambda_{k})>1$.
\begin{align}
\bigg\{P(A+c\lambda_{k}FC)\bigg\}^{s}<-\epsilon E_{n},\quad
k=2,3,\cdots,m
\end{align}
Therefore, we can give following result.
\begin{mycol}
Suppose (A,C) is detectable. $\lambda_{2},\lambda_{3},\cdots,\lambda_{l}$ be the non-zero
eigenvalues of the coupling matrix $L$. If for all $k=2,3,\cdots,m$, $cRe(\lambda_{k})>1$.
Then, the model
\begin{align}
\frac{d x^{i}(t)}{dt}=Ax^{i}(t)
+c\sum\limits_{j=1}^{N}l_{ij}P^{-1}C^{T}C x^{j}(t),\quad i=1,\cdots,N
\end{align}
can be synchronized exponentially, i.e. $x^{i}(t)-\bar{x}(t)\rightarrow 0$ exponentially.
\end{mycol}

It was also given in \cite{Li}. Here, we reveal the relations between \cite{Li} and \cite{Lu2006}.

Based on stabilizable and detectable theory for linear systems, in \cite{Li}, authors discussed following consensus of multiagent systems and
synchronization of complex networks
\begin{eqnarray}
\dot{x}_{i}(t)=Ax_{i}(t)+Bu_{i}(t),~~y_{i}(t)=Cx_{i}(t)
\end{eqnarray}
where $x_{i}(t)\in R^{n}$ is the stat, $u_{i}(t)\in R^{p}$ is the
control input, and $y_{i}(t)\in R^{q}$ is the measured output. $A\in R^{n\times n}$, $B\in R^{n\times p}$, $C\in R^{q\times n}$. It is assumed that is stabilizable and detectable.

An observer-type consensus protocol
\begin{eqnarray}
\dot{v}_{i}(t)=(A+BK)v_{i}(t)+F(\sum_{j=1}^{N}Cl_{ij}v_{j}(t)-\zeta_{i}(t))
\end{eqnarray}
is proposed, which can also be written as
\begin{equation}\label{Li}
\left\{ \begin{array}{ll}
\dot{v}_{i}(t)=(A+BK)v_{i}(t)+\sum_{j=1}^{N}FCl_{ij}(v_{j}(t)-x_{j}(t)) \\
\dot{x}_{i}(t)=Ax_{i}(t)+BKv_{i}(t)
\end{array}
\right.
\end{equation}
where $K\in R^{p\times n}$, $F\in R^{n\times q}$.

Let $e_{i}(t)=v_{i}(t)-x_{i}(t)$, one can transfer (\ref{Li}) to
\begin{equation}\label{e}
\left\{ \begin{array}{ll}
\dot{e}_{i}(t)=Ae_{i}(t)
+FC\sum_{j=1}^{N}l_{ij}e_{j}(t) \\
\dot{x}_{i}(t)=(A+BK)x_{i}(t)+BKe_{i}(t)
\end{array}
\right.
\end{equation}

Denote $\bar{x}(t)=\sum_{i=1}^{N}\xi_{i}{x}_{i}(t)$, $\bar{e}(t)=\sum_{i=1}^{N}\xi_{i}{e}_{i}(t)$,
$\delta x_{i}(t)=x_{i}(t)-\bar{x}(t)$, $\delta\bar{e}_{i}(t)=e_{i}(t)-\bar{e}(t)$.

Therefore, as a special case of Theorem 1, we have
\begin{mythm}\quad
Let $\lambda_{2},\lambda_{3},\cdots,\lambda_{l}$ be the non-zero
eigenvalues of the coupling matrix $L$. If
\begin{eqnarray}
\frac{d z(t)}{dt}=[A+\lambda_{k}FC]z(t),\quad
k=2,3,\cdots,l\label{cs}
\end{eqnarray}
are exponentially stable, then $\delta\bar{e}_{i}(t)=e_{i}(t)-\bar{e}(t)$, $i=1,\cdots,N$, converge to zero exponentially.

Additionally, if $A+BK$ is Hurwiz, then, $\delta\bar{x}_{i}(t)=x_{i}(t)-\bar{x}(t)$, $i=1,\cdots,N$, converge to zero exponentially.

In particular, if $(A,C)$ is detectable, we can pick $F=P^{-1}C^{T}$ and for all $k=2,3,\cdots,m$, $cRe(\lambda_{k})>1$. Then, the model (\ref{Li}) converges.
\end{mythm}

\begin{myrem}

By Theorem 1, it is clear that under the conditions that $A+\lambda_{k}FC,\quad k=2,3,\cdots,l$ are Hurwiz, then
following algorithm discussed in \cite{S,T}
\begin{eqnarray}\label{chen}
\dot{x}_{i}(t)=Ax_{i}(t)+F\sum_{j=1}^{N}l_{ij}y_{j}(t)
\end{eqnarray}
where $y(t)=Cx(t)$, reaches consensus exponentially.

\end{myrem}

\subsection{Applications to Pinning Control}

In this section, we apply general results given in to pinning control of multi-agents consensus.

Consider
\begin{eqnarray}
\left\{\begin{array}{cc}\frac{dx_1(t)}{dt}&=Ax_1(t)
+c\sum\limits_{j=1}^ml_{1j}\Gamma x_j(t)\\
&-c\varepsilon(x_{1}(t)-s(t)),\\
\frac{dx_i(t)}{dt}&=Ax_i(t)+c\sum\limits_{j=1}^m l_{ij}\Gamma x_j(t),\\
&i=2,\cdots,m\end{array}\right.
\end{eqnarray}
In particular, in case $(A,B)$ is controllable,
\begin{eqnarray}
\left\{\begin{array}{cc}\frac{dx_1(t)}{dt}&=Ax_1(t)
+c\sum\limits_{j=1}^m l_{1j}P^{-1}BB^{T}x_j(t)\\
&-c\varepsilon(x_{1}(t)-s(t)),\\
\frac{dx_i(t)}{dt}&=Ax_i(t)+c\sum\limits_{j=1}^ml_{ij}P^{-1}BB^{T}x_j(t),\\
&i=2,\cdots,m\end{array}\right.
\end{eqnarray}
where $\dot{s}(t)=As(t)$.

\begin{proposition} If  $A=(a_{ij})_{i,j=1}^{m}$ is an
irreducible Mezler matrix with $Rank(A)=m-1$. Then, the real part of all
eigenvalues of the matrix
\[\tilde{L}=\left(\begin{array}{cccc}l_{11}-\varepsilon&l_{12}&\cdots&l_{1m}\\
l_{21}&l_{22}&\cdots&l_{2m}\\\vdots&\vdots&\ddots&\vdots\\
l_{m1}&l_{m2}&\cdots&l_{mm}\end{array}\right)\] are negative.
\end{proposition}

Denote $\delta x_i(t)=x_i(t)-s(t)$, then for $i=1,\cdots,m
$, we have
\begin{align}\label{pin1}
\frac{\delta dx_i(t)}{dt}&=Ax_i(t)+c\sum\limits_{j=1}^m \tilde{l}_{ij}\Gamma\delta x_j(t)
\end{align}
and
\begin{eqnarray}\label{pin2}
\frac{d\delta x_i(t)}{dt}=A\delta x_i(t)+c\sum\limits_{j=1}^m\tilde{l}_{ij}P^{-1}BB^{T}\delta x_j(t),
\end{eqnarray}
Denote the eigenvalues of $\tilde{L}$ by $\mu_{1},\cdots,\mu_{N}$ and by same arguments, just replacing $\bar{x}(t)$ by $s(t)$, we have
\begin{mycol}\quad
Let $\mu_{1},\mu_{2},\cdots,\mu_{m}$ be the 
eigenvalues of the coupling matrix $\tilde{L}$. If all variational
equations
\begin{align}
\frac{d z(t)}{dt}=[D f(\bar{x}(t),t)+c\lambda_{k}\Gamma]z(t),\quad
k=2,3,\cdots,l\label{model1a}
\end{align}
are exponentially stable, then for the model $(\ref{pin1})$, $\delta x_{i}(t)=x_{i}(t)-s(t)\rightarrow 0$ exponentially.
\end{mycol}
\begin{mycol}
Suppose (A,B) is controllable. $\mu_{1},\mu_{2},\cdots,\mu_{m}$ be the 
eigenvalues of the coupling matrix $\tilde{L}$. If for all $k=1,2,3,\cdots,m$, $cRe(\mu_{k})>1$.
Then, for the model $(\ref{pin2})$,
$x^{i}(t)-s(t)\rightarrow 0$ exponentially.
\end{mycol}
\begin{myrem}
Synchronization (consensus) with or without pinning control are two different topics but closely related. For Synchronization (consensus) without pinning control, the synchronization state is $\bar{x}(t)$. Instead, For Synchronization (consensus) pinning control, the synchronization state is a solution $s(t)$ of the uncoupled system $\dot{s}(t)=f(s(t))$.
\end{myrem}

\begin{myrem}
For Synchronization (consensus) without pinning control, the coupling matrix $L$ is a singular M-matrix. Instead, For Synchronization (consensus) pinning control, the coupling matrix $L$ is a nonsingular M-matrix..
\end{myrem}

In \cite{Lu2006}, it is revealed that even though
the synchronization manifold can be stable, the individual
state may be unstable. It was also explored that the right and left
eigenvectors of the coupling matrix corresponding to the
eigenvalue 0 play key roles in the geometrical analysis of the
synchronization manifold. These two eigenvectors are used
to decompose the whole space into a direct sum of the
synchronization manifold and the transverse space. By means
of this geometrical analysis, a new approach to
investigating the stability of the synchronization manifold was proposed.

\section{Discussions}
\begin{itemize}
\item
In \cite{Pe}, the synchronization stability of a network of oscillators
by using the master stability function method was introduced.

In \cite{Li}, it was said that \cite{Lu2006,Pe} (References [22] and [27] in \cite{Li})
addressed the synchronization stability of a network of oscillators
by using the master stability function method.

The authors also said that the proposed framework is, in essence,
consistent with the master stability function method used in
the synchronization of complex networks and yet presents a
unified viewpoint to both the consensus of multiagent systems
and the synchronization of complex networks.

In fact, for linear systems, global stability and local stability are equivalent. Therefore, the master stability function method can be used to prove local stability as well as global stability.

It should be pointed out that the master stability functions are different in the two papers \cite{Lu2006} and \cite{Pe}. In \cite{Lu2006}, master stability function applies based on $\bar{x}(t)$. Instead, in \cite{Pe}, master stability function applies  based on $s(t)$ satisfying  $\dot{s}(t)=f(s(t))$. Here, in \cite{Li}, the authors follow the line and approach proposed in \cite{Lu2006}.

\item
There are two fundamental questions about
the synchronization and consensus problems of coupled systems: how to reach
consensus and consensus on what, as said in \cite{Li}.

In fact, this issue has been addressed in \cite{Lu2006} (see Theorem 1 and Theorem 2 in \cite{Lu2006}).
In \cite{Lu2006}, the following universal approach based on the decomposition has been proposed.
\begin{itemize}
\item Synchronization Manifold :
$\mathcal S=\{x=[x^{1},\cdots,x^{N}]\in R^{n,N}: x^{i}=x^{j},~for~all~i,j\} $
  \item Non-orthogonal transverse subspace
  $\mathcal L=\{x=[x^{1},\cdots,x^{N}]\in
    R^{n,N}:\sum\limits_{i=1}^{n}\xi_{i}x^{i}=0\}$
    \item
    Decomposition of $R^{n,N}$:
    $R^{n,N}=\mathcal S\oplus \mathcal L
    $.
    \item For each $x=[x^{1},\cdots,x^{N}]\in
    R^{n,N}$, define
    \begin{itemize}
    \item [1. ] $\bar{x}=\sum\limits_{j=1}^{n}\xi_{i}x^{i}$
    and $\bar{X}=[\bar{x},\cdots,\bar{x}]\in \mathcal S$
    \item [2. ] Let $\delta x^{i}=x^{i}-\bar{x}$, for all $i$.
    Then $\delta x=x-\bar{X}=[\delta x^{1},\cdots,\delta
    x^{m}]\in\mathcal L$,
    \end{itemize}
    \item
    Decomposition: $x=\bar{X}+\delta x$, where $\bar{X}\in \mathcal S$ and
    $\delta x\in\mathcal L$
    \item
    Stability of Synchronization manifold $\mathcal S$
    $\Leftrightarrow$ $\delta x\rightarrow 0.$
    \item
    {\bf $\delta x\rightarrow 0$ answers the question "How to". $\bar{x}=\sum\limits_{j=1}^{n}\xi_{i}x^{i}$ answers "consensus on what", i.e., what is the synchronization state.}
    \item
    {\bf The conditions given in Theorem 1 (as well as Theorem 2) ensure $\delta x\rightarrow 0.$}

\end{itemize}

It has been revealed that the Left eigenvector and Right
Eigenvector of the coupling matrix $L$ with Eigenvalue $0$ Play key roles on Synchronization

\begin{itemize}
    \item  Right eigenvector $\mathbf 1=[1,\cdots,1]^{T}\in
R^{N}$ denotes the direction parallel to $\mathcal S$;
    \item  Left eigenvector $\xi=[\xi_{1},\xi_{2},\cdots,\xi_{N}]^{\top}\in
R^{N}$ denotes
    the direction of the transverse subspace $\mathcal L.$

\end{itemize}

\begin{figure}
\centering\includegraphics[width=.4\textwidth]{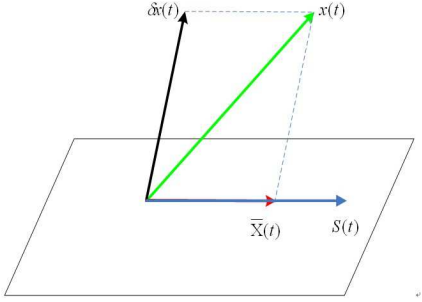}
\caption{Decomposition of $x(t)$} \label{div1}
\end{figure}

\item
In \cite{Li}, the so called relative-State consensus protocol (also see \cite{S,T})
\begin{eqnarray}\label{relative1}
\frac{d x^{i}(t)}{dt}=Ax^{i}(t)
+BL\sum\limits_{j=1}^{N}c\mathcal{L}_{ij} x^{j}(t),\quad i=1,\cdots,N,
\end{eqnarray}
where $L\in R^{p\times n}$, was discussed. \newline

As a direct consequence of Theorem 1, we have
\begin{mythm}\quad
Let $\lambda_{2},\lambda_{3},\cdots,\lambda_{l}$ be the non-zero
eigenvalues of the coupling matrix $\mathcal{L}$. If
\begin{eqnarray}
\frac{d z(t)}{dt}=[A+\lambda_{k}BL]z(t),~
k=2,3,\cdots,l
\end{eqnarray}
are exponentially stable, then the synchronization manifold
$\mathcal S$ is exponentially stable for the general model (\ref{relative1}).
\end{mythm}

On the other hand, in case that the relative states between
neighboring agents are not available, following protocol
\begin{eqnarray}\label{relative2}
\frac{d x^{i}(t)}{dt}=Ax^{i}(t)
+B\bar{L}\sum\limits_{j=1}^{N}c\mathcal{L}_{ij} y^{j}(t),~ i=1,\cdots,N,
\end{eqnarray}
where $\bar{L}\in R^{p\times q}$, $B\in R^{n\times p}$, can be used.

It can also be rewritten as
\begin{eqnarray}\label{relative3}
\frac{d x^{i}(t)}{dt}=Ax^{i}(t)
+B\bar{L}C\sum\limits_{j=1}^{N}c\mathcal{L}_{ij} x^{j}(t),~i=1,\cdots,N,
\end{eqnarray}
Therefore, by Theorem 1, we have
\begin{mythm}\quad
Let $\lambda_{2},\lambda_{3},\cdots,\lambda_{l}$ be the non-zero
eigenvalues of the coupling matrix $L$. If
\begin{eqnarray}
\frac{d z(t)}{dt}=[A+\lambda_{k}B\bar{L}C]z(t),\quad
k=2,3,\cdots,l
\end{eqnarray}
are exponentially stable, then the synchronization manifold
$\mathcal S$ is exponentially stable for the general model (\ref{relative2}).
\end{mythm}

\item
In \cite{Li}, following Spacecraft Formation Flying model
\begin{align}\label{fly}
\left[\begin{array}{c}
\dot{r}_{i}\\
\ddot{r}_{i}\end{array}\right]
&=\left[\begin{array}{cc}
0&I_{3}\\
A_{1}&A_{2}\end{array}\right]
\left[\begin{array}{c}
r_{i}-h_{i}\\
\dot{r}\end{array}\right]\nonumber\\
&+c\sum_{j=1}^{N}a_{ij}\left[\begin{array}{cc}
0&0\\
F_{1}&F_{2}\end{array}\right]
\left[\begin{array}{c}
r_{i}-h_{i}-r_{j}+h_{j}\\
\dot{r}_{i}-\dot{r}_{j}\end{array}\right]
\end{align}
was discussed. And following result (Corollary 3) was given:

Assume that graph has a directed spanning
tree. Then, protocol (\ref{fly}) solves the formation flying problem
if and only if the matrices $\left[\begin{array}{cc}
0&I_{3}\\
A_{1}&A_{2}\end{array}\right]+c\lambda_{i}\left[\begin{array}{cc}
0&0\\
F_{1}&F_{2}\end{array}\right]$are
Hurwitz for $i=2,\cdots,N$, where $\lambda_{i}$, $i=2,\cdots,N$,  denote the
nonzero eigenvalues of the Laplacian matrix of $\mathcal{L}.$

It is clear that Corollary 1, Corollary 2 and Corollary 3 in \cite{Li} can be obtained directly from Theorem 1.
\item
It is claimed in \cite{Li} that "It is observed by comparing Theorem 2 and Corollary 2 that
even if the consensus protocol takes the dynamic form (3) or
the static form (22), the final consensus value reached by the
agents will be the same, which relies only on the communication
topology, the initial states, and the agent dynamics."

However, for the coupled system (4), we have
\begin{eqnarray}
x_{i}(t)-e^{At}\sum_{i=1}^{N}\xi_{i}[x_{i}(0)-v_{i}(0)]\rightarrow 0
\end{eqnarray}
Instead, for the system (16), we have
\begin{eqnarray}
x_{i}(t)-e^{At}\sum_{i=1}^{N}\xi_{i}x_{i}(0)\rightarrow 0
\end{eqnarray}
They are different.

It is claimed in \cite{Li} that Similar to [23], $\delta$ is referred to as the disagreement vector. In fact, the accurate saying is that it came from \cite{Lu2004} and \cite{Lu2006}. In particular, from \cite{Lu2006}.

 \end{itemize}

{\bf Conclusions}
In this paper, we focus on the topic Synchronization and consensus of Complex Networks and their relationships. It is revealed that two topics are closely relating to each other and all results given in \cite{Li} can be obtained by the results in \cite{Lu2006}.
Several protocols on this topic are also revisited and the relationships between them are addressed. It is pointed out that the model introduced in \cite{Lu2006} and the approach provided there is universal. Many existed synchronization and consensus models and their stability behavior analysis can be derived easily from the theoretical results given in \cite{Lu2006}. These models include consensus and synchronization of linear coupled nonlinear (or linear) systems, observed-based linear systems and many others.

\end{document}